\documentclass[epj]{svjour}

\usepackage{amsmath}
\usepackage{amssymb}
\usepackage{graphicx}

\newcommand{\bs}{\boldsymbol}

\newcommand{\mr}{\mathrm}

\begin{document}

\title{A nonadiabatic semi-classical method for dynamics of atoms in optical lattices}

\author{S.\ Jonsell, C.\ M.\ Dion, M.\ Nyl\'en, S.\ J.\ H.\ Petra,
  P.\ Sj\"olund, A.\ Kastberg} 

\institute{Department of Physics, Ume\aa\ University, SE-901 87,
  Ume\aa, Sweden, \email{jonsell@tp.umu.se}.}

\abstract{We develop a semi-classical method to simulate the motion of
atoms in a dissipative optical lattice. Our method treats the
internal states of the atom quantum mechanically, including all
nonadiabatic couplings, while position and momentum are treated as
classical variables. We test our method in the one-dimensional
case. Excellent agreement with fully quantum mechanical simulations is
found. Our results are much more accurate than those of earlier
semi-classical methods based on the adiabatic approximation.}

\PACS{
{32.80.Pj}{Optical cooling of atoms; trapping}\and
{03.65.Sq}{Semiclassical theories and applications}
}

\authorrunning{S. Jonsell et al.}

\maketitle

\section{Introduction}


One of the most spectacular achievements in the field of laser cooling
is the discovery of cooling below the Doppler limit in optical
lattices, so called Sisyphus cooling \cite{let88}. An optical lattice
is a standing wave of laser light, forming a periodic light-shift
potential for atoms moving in the laser field \cite{jes96,gry01}.  In
the optical lattices used for cooling the frequency of the lasers are
tuned {\it close to an atomic resonance}. The atoms thus undergo
cycles of absorption followed by spontaneous emission.  Under the
right experimental conditions, the spontaneous emission causes an
overall loss of kinetic energy of the atoms, i.e., cooling.

Optical lattices are also widely used in Bose-Einstein condensation
experiments \cite{blo05} and for quantum state manipulation
\cite{mon02}. These lattices are tuned {\it far from atomic
  resonances}, in order to avoid spontaneous emission which would
destroy the coherence of the condensate. Therefore these far detuned
lattices do not provide any cooling.

The name Sisyphus cooling comes from the first theoretical model for
the process \cite{dal89,ung89}. This model is based on optical pumping
between the magnetic sublevels of the light shifted atomic ground
state.
However, at least in its original form it relies on a number of
simplifying assumptions, such as a semi-classical approximation,
spatial averaging, and a simplified level structure (a ground state
with angular momentum $J_{\mr g}=1/2$, and an excited state with
angular momentum $J_{\mr e}=3/2$). Whereas this model correctly
predicts some qualitative features of cooling in optical lattices, it
is too crude to provide an overall quantitative agreement. Instead, a
number of more advanced theoretical methods have been developed. The
most accurate of these is the Monte-Carlo wavefunction technique
\cite{cas95}, a fully quantum mechanical method based on stochastic
wavefunctions.

In this paper, we develop and test a new semi-classical method for
simulating the motion of atoms in a near-resonant optical lattice. 
The most important approximation of our method is that
the position and momentum of the atoms are treated as classical
variables. Other approximations include a classical treatment of the
light field, and adiabatic elimination of excited states of the atoms,
but otherwise we make as few approximations as possible. In
particular, the internal states are treated quantum mechanically,
allowing for any kind of coherent superposition between magnetic
sublevels.

Even though more exact fully quantum mechanical theoretical methods
exist, semi-classical methods are valuable, partly because they are
less demanding numerically, but also because they provide a simpler
conceptual framework in which it is easier to formulate an intuitive
picture of e.g.\ the mechanisms involved in the cooling process.
Up to now, all semi-classical methods for laser cooling in optical
lattices have been based on atoms that are pumped between definite
internal states as they move through the lattice.  To this end a basis
of so-called {\it adiabatic} states, diagonalizing the light-shift
potential at every position, has been used instead of the {\it
  diabatic} basis of the magnetic substates \cite{pet99}.  Coherences
between adiabatic states have not been included in the description,
and neither have so-called nonadiabatic couplings arising from the
position dependence of the adiabatic basis. Thus, the motion of the
atoms is described by purely classical equations, albeit the various
potentials, pumping rates and diffusion coefficients have been derived
from a quantum-mechanical origin. These adiabatic semi-classical
methods reproduce some of the qualitative features of Sisyphus
cooling, e.g.\ a linear relation between temperature and irradiance at
high irradiances \cite{jer03}. However, we show that even at very high irradiances
the slope of this linear dependence does not agree with fully
quantum-mechanical simulations.  At the lower irradiances relevant to
most experiments the adiabatic semi-classical method deviates even
more severely from the fully quantum-mechanical results. Both these
problems are solved by the nonadiabatic semi-classical approach.

\section{Theory}

In this section we develop the basic semi-classical equations of
motion, on which our simulations are based. For generality the theory
is developed in three dimensions.  The angular momenta of the ground
and excited states of the lattice transition are denoted by $J_{\mr
  g}$ and $J_{\mr e}$ respectively, and the corresponding magnetic
quantum numbers are $M_{\mr g}$ and $M_{\mr e}$.  Although the
derivation is more general, we shall in the end apply the theory to
the case $J_{\mr e}=J_{\mr g}+1$.  Also, the light field
${\bs\xi}({\bs r})$ creating the lattice could take different forms,
but will in the end be assumed to have a lin$\perp$lin configuration
in one, two or three dimensions \cite{gry01}. That is, the lattice is
created by the interference pattern of light fields, forming lattice
sites with alternating $\sigma^+$ and $\sigma^-$ polarizations.

We start from the optical Bloch equations for an atom in a standing
wave laser field \cite{coh90}.  They can be derived under very general
conditions, and represent for practical purposes an exact fully
quantum mechanical description of atomic motion in an optical lattice.
Our first important approximation is that the population of the
excited state is sufficiently low to allow its adiabatic elimination.
The condition for this is that the saturation parameter
\begin{equation}
s_0 = \frac{\Omega^2/2}{\Delta^2+\Gamma^2/4}\ll 1.
\end{equation}
Here $\Delta$ is the detuning from resonance, $\Gamma$ the natural
width of the excited state, and $\Omega$ is the Rabi
frequency\footnote{We use the Rabi frequency based on the {\it total}
  laser field. This is the same convention as was used, e.g., in Ref.\ 
  \cite{pet99}. Sometimes the Rabi frequency is instead on the laser
  irradiance {\it per beam}, which for a one-dimensional lin$\perp$lin
  configuration is half the total irradiance.}.  The details of the
adiabatic elimination of the excited states can be found e.g.\ in
Ref.\ \cite{coh90}. This approximation is an important simplification,
since it reduces the master equation for the full density matrix, to
an equation for the $(2J_{\mr g}+1)\times(2J_{\mr g}+1)$ density
matrix $\sigma$ of the ground states. The resulting equation for the
evolution of $\sigma$ reads
\begin{equation}
  \label{eq:master}
  {\mr i}\hbar \dot\sigma=\left[\hat{\rm H},\sigma\right]+
{\mr i}\hbar
\left.\dot  \sigma\right|_{\text{sp}}.
\end{equation}
The first term on the right-hand side of this equation represents the
Hamiltonian part of the evolution. The second term represents the
non-Hermitian evolution due to spontaneous emission.  The Hamiltonian
contains the kinetic term and the light-shift potential,
\begin{equation}
  \label{eq:Ham}
  \hat{\rm H}=\frac{\hat{\bf p}^2}{2m}+\hbar\Delta'\hat{\mr A}({\bs r}),
\end{equation}
where $\hat{\bf p}$ is the momentum operator of the atom, ${\bs r}$
its position, $\Delta'=\Delta s_0/2$, and the operator $\hat{\mr
  A}({\bs r})$ is given by
\begin{equation}
  \label{eq:A}
  \hat{\mr A}({\bs r})=\left[\hat{\mathbf d}^- \cdot {\bs \xi}^*({\bs r})\right]
\left[\hat{\mathbf d}^+ \cdot {\bs \xi}({\bs r})\right].
\end{equation}
Here $\hat{\mathbf d}^+$ is an operator that promotes an atom from the
ground to the excited state, while $\hat{\mathbf d}^-=(\hat{\mathbf
  d}^+)^\dagger$ is responsible for the reverse process.  In the basis
of circular polarization vectors
\begin{equation}
  \label{eq:polvec}
  \hat{\bs\varepsilon}_{\pm1}=\mp\frac{1}{\sqrt{2}}\left(\hat{\mathbf x}\pm
    \hat{\mathbf y}\right),\quad \hat{{\bs\varepsilon}}_0=\hat{\mathbf z},
\end{equation}
they have simple expressions in terms of Clebsch-Gordan coefficients
\begin{equation}
  \label{eq:ddef}
  \hat{d}^+_q  =\langle J_{\mr e} M_{\mr e} \mid J_{\mr g} 1 M_{\mr g} q\rangle=   \left(\hat{d}^-_q\right)^*.
\end{equation}
In the basis of the magnetic substates $M_g$ the operator $\hat{\mr
  A}({\bs r})$ is represented a matrix $A({\bs r})$.

For the simple model atom with $J_{\mr g}=1/2$ and $J_{\mr e}=3/2$
$A({\bs r})$ is a diagonal matrix. However, most atoms of interest
have a more complicated level structure, including non-diagonal
couplings in the potential. Therefore previous semi-classical methods
have used an {\it adiabatic} basis, where the atomic states are the
eigenstates of $A({\bs r})$. Whereas $A({\bs r})$ is diagonal in the
adiabatic basis, the position dependence of the basis gives rise to
nonadiabatic couplings between adiabatic states.  In the
adiabatic approximation these couplings are neglected.
In our method we keep all off-diagonal couplings. The results are then
independent of the basis used, and the simplest choice is to stay with
the magnetic levels $M_{\mr g}$, the  {\it diabatic} basis.
Since this basis is the same for all ${\bs r}$, all couplings are
included in $A({\bs r})$, and their functional form can be calculated
analytically for a given laser configuration.

The second term on the right-hand side of Eq.\ (\ref{eq:master})
contains processes associated with spontaneous emission.  Writing the
matrix elements of $\sigma$ in the position representation, $\langle
{\bs r} \mid \sigma\mid {\bs r}'\rangle=\sigma({\bs r},{\bs r}')$, its
form is
\begin{align}
  \label{eq:spont}
  \left.\dot \sigma({\bs r},{\bs r}')\right|_{\text{sp}}
  =&-\frac{\Gamma'}{2}\left[A({\bs r})\sigma({\bs r},{\bs r}')
    +\sigma({\bs r},{\bs r}')A({\bs r}')\right] +
  \frac{3\Gamma'}{8\pi} \nonumber\\&\times \int {\rm d}\Omega_{{\bs
      k}} \sum_{\bs\epsilon\perp \bs k} B_{\bs\epsilon}^\dagger({\bs
    r}){\mr e}^{-{\mr i}{\bs k}\cdot{\bs r}}\sigma({\bs r},{\bs r}')
  {\mr e}^{{\mr i}{\bs k}\cdot {\bs r}'}B_{\bs \epsilon}({\bs r'}),
\end{align}
where $\Gamma'=\Gamma s_0/2$.  The matrices $B_{\bs \epsilon}({\bs
  r})$ are given by
\begin{equation}
  \label{eq:B}
  B_{\bs \epsilon}({\bs r})=\left[\hat{\mathbf d}^- \cdot {\bs \xi}^*({\bs r})\right]
\left[\hat{\mathbf d}^+\cdot {\bs \epsilon}\right].
\end{equation}
Hence, $B^\dagger_{\bs \epsilon}$ represents the excitation of an atom
by the laser field, and its subsequent return to the ground state via
spontaneous emission of a photon with polarization $\bs\epsilon$. The
factors $\exp({\mr i}{\bs k}\cdot{\bs r})$ represent the atomic recoil from
a spontaneously emitted photon with wave vector ${\bs k}$. The
integration is over the directions of the emitted photon, and the
summation is over any basis spanning the allowed polarization vectors.
The recoil momentum of the atomic transition is $p_{\mr R}=\hbar
k_{\mr R}=\hbar |{\bs k}|$.

Our goal is to approximate Eq.\ (\ref{eq:master}) by a semi-classical
equation where every atom has a definite position and momentum, i.e.\ 
every atom follows a trajectory in phase space. This is of course not
allowed in quantum mechanics, because of the uncertainty principle.
Hence, a quantum mechanical phase space cannot be defined, but it is
still possible to introduce a ``coarse grained'' version of phase
space through the Wigner function
\begin{equation}
  \label{eq:wigner}
  W({\bs r},{\bs p},t)=\frac{1}{h^3}\int {\rm d}{\bs u}\left\langle {\bs
      r}+\frac{\bs u}{2} \right| \sigma \left|  {\bs r}-\frac{\bs u}{2}
\right\rangle {\mr e}^{-{\mr i} {\bs p}\cdot {\bs u}/\hbar}.
\end{equation}
In this work the Wigner function is a matrix with dimension $2J_{\mr
  g}+1$.  The Wigner transformation of Eq.\ (\ref{eq:master}) becomes
\begin{align} \nonumber
\label{eq:wigner2}
\left( \frac{\partial}{\partial t} +\frac{{\bs p}}{m}  \cdot\nabla_{\bs
    r}\right)&  W({\bs r},{\bs p},t)=\\ \nonumber & {\mr
  i}\frac{\Delta'}{\hbar^3} \int {\mr d}{\bs q} {\mr e}^{{\mr i}{\bs
    q}\cdot{\bs r}/\hbar} \big[W({\bs r},{\bs p}+\frac{{\bs
    q}}{2},t)\tilde{A}({\bs q}) \\ \nonumber & \qquad\qquad\qquad -
\tilde{A}({\bs q}) W({\bs r},{\bs p}-\frac{{\bs q}}{2},t) \big]\\
\nonumber - & \frac{\Gamma'}{2\hbar^3} \int {\mr d}{\bs q} {\mr
  e}^{{\mr i}{\bs q}\cdot{\bs r}/\hbar} \big[W({\bs r},{\bs
  p}+\frac{{\bs q}}{2},t)\tilde{A}({\bs q}) \\ \nonumber &
\qquad\qquad\qquad +
\tilde{A}({\bs q}) W({\bs r},{\bs p}-\frac{{\bs q}}{2},t) \big]\\
\nonumber + & \frac{3\Gamma'}{8\pi \hbar^6}\int {\mr d}\Omega_{\bs k}
\sum_{{\bs \epsilon}\perp{\bs k}}\int {\mr d}{\bs q}\int {\mr d}{\bs
  q}' {\mr e}^{{\mr i}({\bs q}-{\bs q}')\cdot{\bs r}/\hbar}
\\
&\times \tilde{B}_{\bs \epsilon}^\dagger({\bs q}') W({\bs r},{\bs
  p}+\hbar{\bs k}+\frac{{\bs q}'+{\bs q}}{2},t) \tilde{B}_{\bs
  \epsilon}({\bs q}).
\end{align}
Here $\tilde{A}$ and ${\tilde B}_{\bs\epsilon}$ are the Fourier
transforms
\begin{align}
\label{eq:Ak}
\tilde{A}({\bs q}) & =\int{\mr d}{\bs r}{\mr e}^{-{\mr i}{\bs q}\cdot
  {\bs r}/\hbar}
A({\bs r}),\\
\label{eq:Bk}
\tilde{B}_{\bs \epsilon}({\bs q}) & = \int {\mr d}{\bs r}{\mr
  e}^{-{\mr i}{\bs q}\cdot {\bs r}/\hbar} B_{\bs \epsilon}({\bs r}).
\end{align}

No approximation has been made in going from Eq.\ (\ref{eq:master}) to
Eq.\ (\ref{eq:wigner2}), the Wigner transformation is just another
representation of the same physics. Now, we introduce the
semi-classical approximation. According to this approximation the
momentum distribution varies smoothly and slowly over typical momentum
transfers ${\bs q}$ in Eq.\ (\ref{eq:wigner2}). Since $A({\bs r})$ and
$B_{\bs \epsilon}({\bs r})$ have the same periodicity as the laser
field, i.e.\ $\lambda=2\pi/k_{\mr R}$, Eqs.\ (\ref{eq:Ak}) and (\ref{eq:Bk})
show that the typical size for ${\bs q}$ is the recoil momentum. Thus,
the semi-classical approximation assumes that the momentum
distribution changes little for emission/absorption of a single
photon. As long as the atomic momenta are several recoil units large,
and the effects of quantization of the atomic states are small, this
approximation can be expected to work well.

Invoking the semi-classical approximation we can make a second-order
Taylor expansion around ${\bs p}$ of the Wigner distribution
\begin{align} \nonumber
  W({\bs r},{\bs p}+{\bs q},t)\simeq& W({\bs r},{\bs p},t) + {\bs
    q}\cdot\nabla_{\bs p} W({\bs r},{\bs p},t) \\ &+\frac{1}{2}({\bs
    q}\cdot \nabla_{\bs p})^2 W({\bs r},{\bs p},t).
  \label{eq:taylor}
\end{align}
Using this expansion it is possible to replace $\tilde{A}$ and
$\tilde{B}_{\bs \epsilon}$ by their counterparts in position space.
The resulting equation for the semi-classical Wigner function, which
now can be interpreted as a phase-space distribution, is
\begin{align} \nonumber
  \bigg(\frac{\partial}{\partial t}+ & \sum_{i=1}^3 \frac{p_i}{m}
  \partial_i \bigg)W({\bs r},{\bs p},t)= {\mr i}\Delta'\left[W({\bs
      r},{\bs p},t),A({\bs r})\right] \\ \nonumber &
  -\frac{\Gamma'}{2}\left\{W({\bs r},{\bs p},t),A({\bs r})\right\}
  \\ \nonumber & + \Gamma'\sum_{q=0,\pm1} B_q^\dagger({\bs r})W({\bs
    r},{\bs p},t)B_q({\bs r}) \\ \nonumber &
  +\frac{\hbar\Delta'}{2}\sum_{i=1}^3\left\{\partial_{p_i} W({\bs
      r},{\bs p},t),\partial_iA({\bs r}) \right\} \\  \nonumber &
  +\frac{{\mr i}\hbar\Gamma'}{4}\sum_{i=1}^3\left[\partial_{p_i}W({\bs
      r},{\bs p},t),\partial_i A({\bs r}) \right] \\ \nonumber & +
  {\mr i}\frac{\hbar\Gamma'}{2}\sum_{q=0,\pm1}\sum_{i=1}^3
  [\partial_iB_q^\dagger({\bs r})\partial_{p_i}W({\bs r},{\bs
    p},t)B_q({\bs r})\\ \nonumber& \qquad\qquad\qquad\qquad\qquad
  -B_q^\dagger({\bs r})\partial_{p_i}W({\bs r},{\bs
    p},t)\partial_iB_q({\bs r})] \\ \nonumber & -{\mr
    i}\frac{\hbar^2\Delta'}{8}\sum_{i=1}^3\sum_{j=1}^3\left[\partial_{p_i}\partial_{p_j}W({\bs
      r},{\bs p},t),\partial_i\partial_jA({\bs r}) \right]
  \\ \nonumber & +\frac{\hbar^2\Gamma'}{16}\sum_{i=1}^3\sum_{j=1}^3
  \left\{\partial_{p_i}\partial_{p_j}W({\bs r},{\bs
      p},t),\partial_i\partial_jA({\bs r}) \right\} \\ \nonumber &
  -\frac{\hbar^2\Gamma'}{8}\sum_{q=0,\pm1}\sum_{i=1}^3\sum_{j=1}^3
  \big[\partial_i\partial_jB_q^\dagger({\bs r})
  \partial_{p_i}\partial_{p_j}W({\bs r},{\bs p},t)B_q({\bs r}) \\
  \nonumber & \qquad\qquad\qquad -2\partial_iB_q^\dagger({\bs r})
  \partial_{p_i}\partial_{p_j}W({\bs r},{\bs p},t)\partial_jB_q({\bs
    r}) \\ & \qquad\qquad\qquad + B_q^\dagger({\bs r})
  \partial_{p_i}\partial_{p_j}W({\bs r},{\bs p},t)
  \partial_i\partial_jB_q({\bs r}) \big] \nonumber \\ &
  +\frac{\hbar^2k_{\mr R}^2\Gamma'}{5}\sum_{q=0,\pm1}\sum_{i=1}^3
  \eta_{i,q} B_q^\dagger({\bs r}) \partial_{p_i}^2W({\bs r},{\bs p},t)
  B_q({\bs r}).
\label{eq:fullsc}
\end{align}
In this equation we use the short-hand notation $\partial_i\equiv
\partial/\partial r_i$, $\partial_{p_i}\equiv\partial/\partial p_i$,
where $i=x,y,z$ are the Cartesian coordinates.  The constants
$\eta_{i,q}$ come from the integration over the direction of the
spontaneously emitted photon, and are given by
\begin{align} \nonumber
  \eta_{x,\pm1} =\eta_{y,\pm1}= 3/4,\qquad &\eta_{z,0} =1/2\\
  \eta_{x,0} =\eta_{y,0}= 1,\qquad &\eta_{z,\pm1} =1.
\end{align}
Although the equation is somewhat lengthy, it is possible to give
physical interpretations to its terms. The left hand side is simply
the kinetic term, i.e.\ the full derivative ${\mr d}/{\mr d}t$. On the
right, the terms where $W({\bs r},{\bs p},t)$ appear without any
derivative represent transfer of population between states, either by
couplings from non-diagonal terms of the light-shift potential
$\Delta'A({\bs r})$, or by optical pumping. The terms containing
$\partial_{p_i}W({\bs r},{\bs p},t)$ describe the motion of the atoms
due to forces from light-shift potential and the radiation pressure.
Terms containing second derivatives of both $W({\bs r},{\bs p},t)$ and
first or second derivatives of
$A({\bs r})$ or $B_{\bs \epsilon}({\bs r})$ describe the momentum
diffusion due to fluctuations in the number of photons absorbed.
Finally, the term containing $\partial_{p_i}^2W({\bs
  r},{\bs p},t)$, but no other derivatives, contains the momentum
diffusion due to the recoil kick from spontaneously emitted photons.

Equation (\ref{eq:fullsc}) is the most complete semi-classical
approximation for the time-dependent distributions of atoms in ${\bs
  r}$ and ${\bs p}$ space. It is classical in the sense that the atoms
are assumed to be particles with definite positions and momenta. The
internal states, however, are treated fully quantum mechanically,
including all off-diagonal couplings and coherences. It is thus not
possible to assign an atom to a definite internal state, nor is it
described as a classical probability distribution over the
different internal states, but as a quantum-mechanical superposition
of internal states.

In order to solve Eq.\ (\ref{eq:fullsc}) we recast it into a
Langevin-type equation. That is, instead of calculating distributions
of atoms, we shall calculate phase-space trajectories $\tilde{\bs
  x}(t)$ and $\tilde{\bs p}(t)$ of individual atoms. In doing this, we
still want to keep the quantum mechanical description of the internal
states. That is, the probability distribution of an atom is
\begin{equation}
  \label{eq:WLangevine}
  W({\bs r},{\bs p},t)=w(t)\delta\left({\bs r}-\tilde{\bs r}(t)\right)
\delta\left({\bs p}-\tilde{\bs p}(t)\right).
\end{equation}
Here $w(t)$ is a matrix of dimension $2J_{\mr g}+1$ containing the
internal-state density matrix of the atom at time $t$. Inserting this
form into Eq.\ (\ref{eq:fullsc}), and integrating over position and
momentum, the evolution equation for $w(t)$ is obtained
\begin{align}
  \label{eq:wdot} \nonumber
  \dot{w}(t) =&{\mr i}\Delta'[w(t),A({\bs
    r})]-\frac{\Gamma'}{2}\{w(t),A({\bs r})\}\\ &
  +\Gamma'\sum_{q=0,\pm1}B_q^\dagger({\bs r})w(t)B_q({\bs r}).
\end{align}
Here and below, we use the simplified notation ${\bs r}$ for
$\tilde{\bs r}(t)$ and ${\bs p}$ for $\tilde{\bs p}(t)$. It is,
however, important to understand that these are now time-dependent
functions representing position and momentum of a single atom, which
are conceptually very different from the variables in Eq.\ 
(\ref{eq:fullsc}).  Using that $\langle {\bs x} \rangle=\text{
  Tr}\{{\bs x} w\}$ etc., we derive the equations for the evolution of
${\bs x}$ and ${\bs p}$ (see, e.g., \cite{risken})
\begin{align}
  \dot{\bs x} & = \frac{\bs p}{m},  \label{eq:xdot}\\
  \dot{\bs p} & = {\bs f}(t)+{\bs \chi}(t). \label{eq:pdot}
\end{align}
Here ${\bs f}(t)$ is a force and ${\bs \chi}(t)$ is a
fluctuating force with the properties
\begin{equation}
  \label{eq:chi}
  \langle \chi_i(t)\rangle=0,\quad \langle \chi_i(t)\chi_j(t')\rangle=2 D_{ij}(t) \delta(t-t').
\end{equation}
The force is given by
\begin{align}
  \label{eq:force}
  f_i(t)=& -\hbar\Delta' \text{Tr}\left\{\partial_iA({\bs
      r})w(t)\right\}
  \nonumber \\
  &-{\mr i}\frac{\Gamma'}{2}\sum_{q=0,\pm1}\text{Tr}\big\{[B_q({\bs
    r})\partial_iB_q^\dagger({\bs r}) \nonumber \\ & \qquad\qquad -
  \partial_i B_q({\bs r})B^\dagger_q({\bs r})]w(t)\big\}.
\end{align}
The first term above is the force arising from the second-order
light-shift potential, while the second term is the radiation
pressure.  The diffusion coefficient is
\begin{align}
  D_{ij}(t) =&\delta_{ij}\frac{\Gamma'\hbar^2k_{\mr
      R}^2}{5}\sum_{q=0,\pm1}\eta_{i,q} \text{Tr}\left\{B_q({\bs
      r})B_q^\dagger({\bs r})w(t)\right\}
  \nonumber \\
   &+\frac{\Gamma'\hbar^2}{2(1+\delta_{ij})}\sum_{q=0,\pm1} \text{
    Tr}\big\{\big[\partial_iB_q({\bs r})\partial_jB_q^\dagger({\bs r})
  \nonumber \\  & \qquad\qquad\qquad +\partial_jB_q({\bs
    r})\partial_iB_q^\dagger({\bs r})\big]w(t)\big\}.
\end{align}
The first term arises from the recoil from photons spontaneously
emitted in random directions, while the second term is connected to
fluctuations in the radiation pressure. The latter term is in general
anisotropic.

\section{Numerical implementation}

We simulate the equations (\ref{eq:wdot}), (\ref{eq:xdot}) and
(\ref{eq:pdot}) in one dimension. The laser field has the form
\begin{equation}
  \label{eq:xi}
{\bs\xi}(z)=\cos(k_{\mr R}z){\bs \varepsilon}_{-1}-{\mr i}\sin(k_{\mr R}z){\bs \varepsilon}_{+1}.
\end{equation}
At the start of every time step the
system is in a pure quantum mechanical state. For every time step
 $w$, $z$, and $p$ are evolved using a second-order Runge-Kutta
method. The fluctuating force $\chi(t)$ is included as a term
\begin{equation}
  \label{eq:diff}
  r\sqrt{2D\, {\mr d}t},
\end{equation}
where $r$ is a random number with zero average and unit variance. This
term only needs  to be evaluated once every time step \cite{hon92}.

At the end of a time step, the system will not be in a pure state
anymore. Its internal-state density matrix $w$ can, however, be
decomposed  into $2J_{\mr g}+1$ pure states
\begin{equation}
  \label{eq:wdecomp}
  w=\sum_{i=1}^{2J_{\mr g}+1}\lambda_i |\Phi_i\rangle\langle\Phi_i|.
\end{equation}
The coefficient $\lambda_i$ are the eigenvalues, and $|\Phi_i\rangle$ the
corresponding eigenvectors, of $w$. Since $w$ is a density matrix, the
eigenvalues satisfy the properties $\lambda_i>0$ and
$\sum_{i=1}^{2J_{\mr g}+1} \lambda_i=1$, 
and can be interpreted as classical probabilities of the different
states $|\Phi_i\rangle$  \cite{peres}.
Hence, at the end of each time step the
system has the probability $\lambda_i$ to make a  ``jump'' into
the pure state $|\Phi_i\rangle$.
Even though a density matrix in general has an infinite
number of decompositions into pure states, the decomposition above is
unique in the sense that it is the only one into a set of linearly
independent pure states.

For numerical efficiency the eigenvalues were obtained
using first-order perturbation theory, which is sufficiently exact if
${\mr d}t$ is short  enough.  In practice, one of the eigenvalues will be
very close to one, while the others are small or zero. Thus, one can
interpret the system as either staying in the  same state, or
jumping to a new state.  When the eigenvalues obtained by perturbation
theory
indicate that the system makes  a jump, the accuracy is increased by a full
diagonalization of $w$. 
The expense in computer time for this improvement is modest, 
since jumps are comparatively rare.

\section{Results}

\begin{figure}[htbp]
   \includegraphics*[width=9cm]{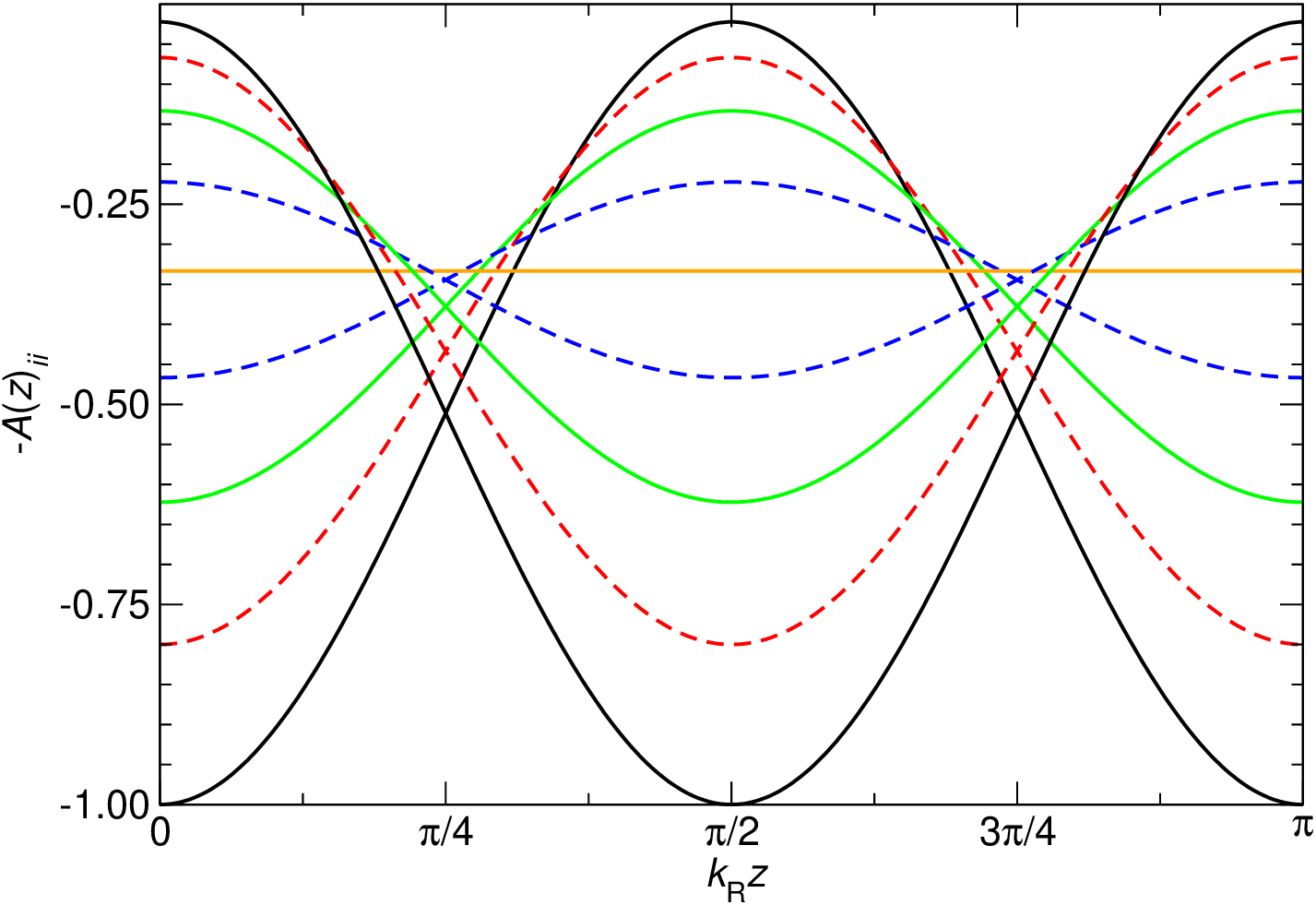}
   \caption{Diagonal elements of the diabatic potential for the
     $J_{\mr g}=4\rightarrow J_{\mr e}=5$ transition. Each curve
     corresponds to a magnetic sublevel $M_{\mr g}$ of the ground
     state. Curves corresponding to $\pm|M_{\mr g}|$ share the same
     color coding, and differ only by the phase $\pi/2$. States with
     $M_{\mr g}$ even (odd) are represented by solid (dashed) curves. }
   \label{fig:potentials}
 \end{figure}

In our simulations we used the parameters for the D2 line in cesium,
i.e. $J_{\mr g}=4$, $J_{\mr e}=5$, and natural width
$\Gamma/2\pi=5.2227$~MHz, and recoil energy $E_{\mr
  R}=1.3692\times10^{-30}$~J \cite{steck}.
The diagonal elements of the diabatic potential for this transition
are displayed in Figure \ref{fig:potentials}.  We first investigated
the steady-state momentum distributions.  For potential depths
$\hbar|\Delta'|\geq 200E_{\mr R}$ the samples contained 5000 atoms,
and were iterated for the time $2500/\Gamma'$. To improve statistics
the momentum distribution was averaged over the last $1000/\Gamma'$ of
the evolution time.  For low potential depths convergence is slower.
Therefore we used 20000 atoms for $\hbar\Delta'<200E_{\mr R}$, and the
evolution time $5000/\Gamma'$, with averaging over the last
$2000/\Gamma'$. For all runs the time step was ${\mr
  d}t=0.025/\Gamma'$, and the initial state a spatially uniform
distribution with temperature of 10 $\mu$K.

\begin{figure}[htbp]
  \includegraphics*[width=9cm]{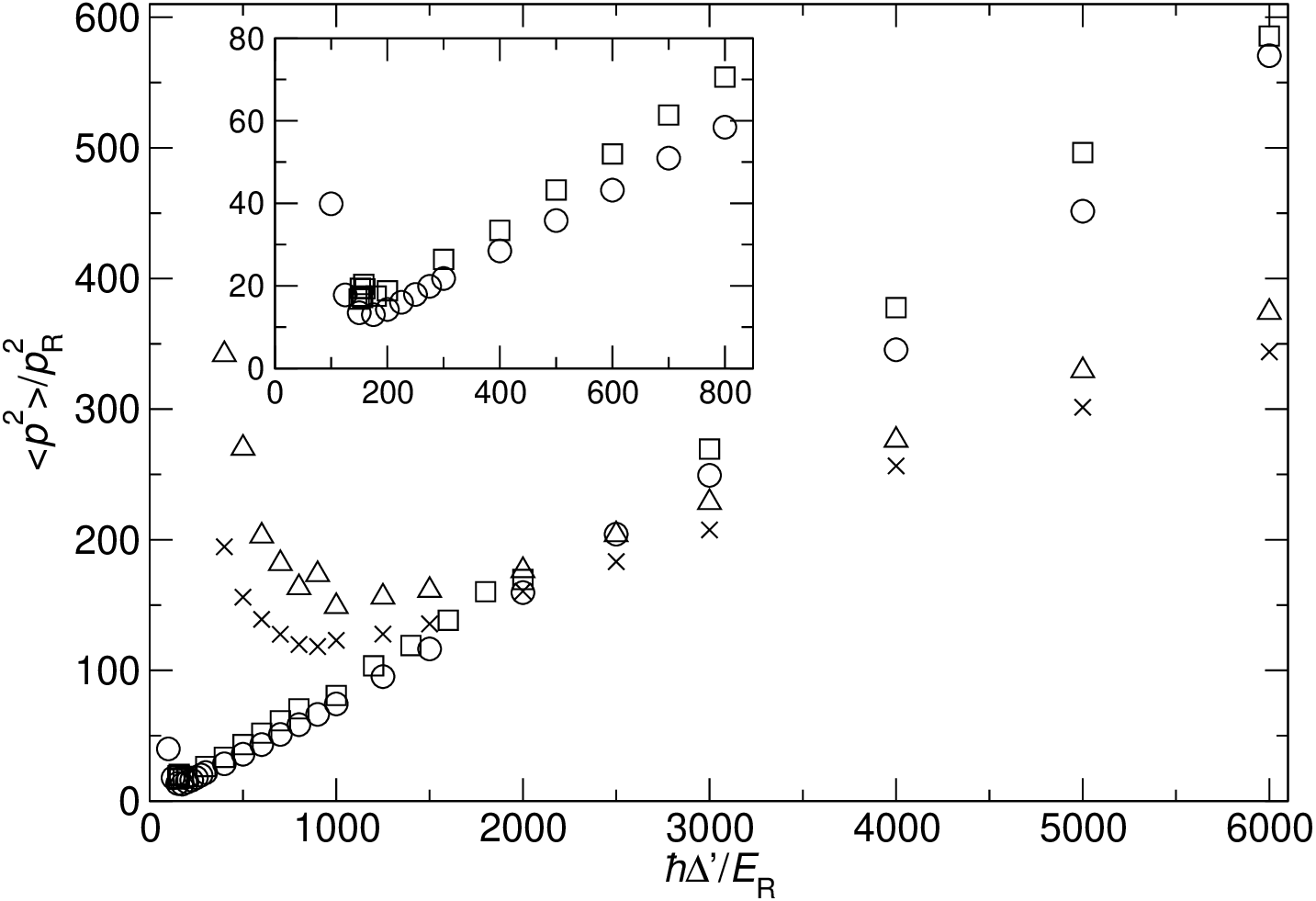}
  \caption{Semi-classical results for $\langle p^2\rangle$ (circles)
    compared to full quantum results (squares). For comparison we also
    show results based on the adiabatic approximation calculated
    similarly to the method used in Ref.\ \protect\cite{pet99}
    (crosses), and the same method improved by including also
    non-diagonal diffusion coefficients (triangles).  The detuning is
    $\Delta=-10\Gamma$.  }
  \label{fig:p2}
\end{figure}

 Results for $\langle p^2\rangle$ as a function of potential depth
 $\hbar\Delta'/E_{\mr R}$ for a detuning $\Delta=-10\Gamma$ are
 displayed in Figure \ref{fig:p2}. Our results are compared to a
 full-quantum simulation using the Monte-Carlo wave function method
 \cite{cas95}. The two methods are in excellent agreement. The
 relative difference is at most about 20\%. It is not clear how much
 of this deviation can be attributed to the fundamental difference
 between the two methods, and how much is due to e.g.\ statistical
 uncertainties or other numerical errors. For deep potentials both
 methods give the same linear slope, although with a slight offset.
 The agreement continues all the way down through {\it d\'ecrochage},
 i.e. the point where the curve turns around and starts to increase
 again for small potential depths, although statistical fluctuations
 in the full-quantum data make comparisons more difficult here.
 
 It also evident from Figure \ref{fig:p2} that the present method is a
 substantial improvement of the adiabatic method used in Ref.\ 
 \cite{pet99}. The methods do not even agree at large potential
 depths, where one would expect the nonadiabatic corrections to become
 small. Improving upon this method by including non-diagonal diffusion
 terms (for details see Ref.\ \cite{pet99}) does not substantially
 change the situation. We note that even in the limit of vanishing
 nonadiabatic corrections our method differs from that in Ref.\ 
 \cite{pet99} by allowing for coherences between the internal states.
 In the adiabatic basis the potential does not induce any coherences
 between internal states, but such coherences are still induced by
 optical pumping.

\begin{figure}[tbp]
  \includegraphics*[width=9cm]{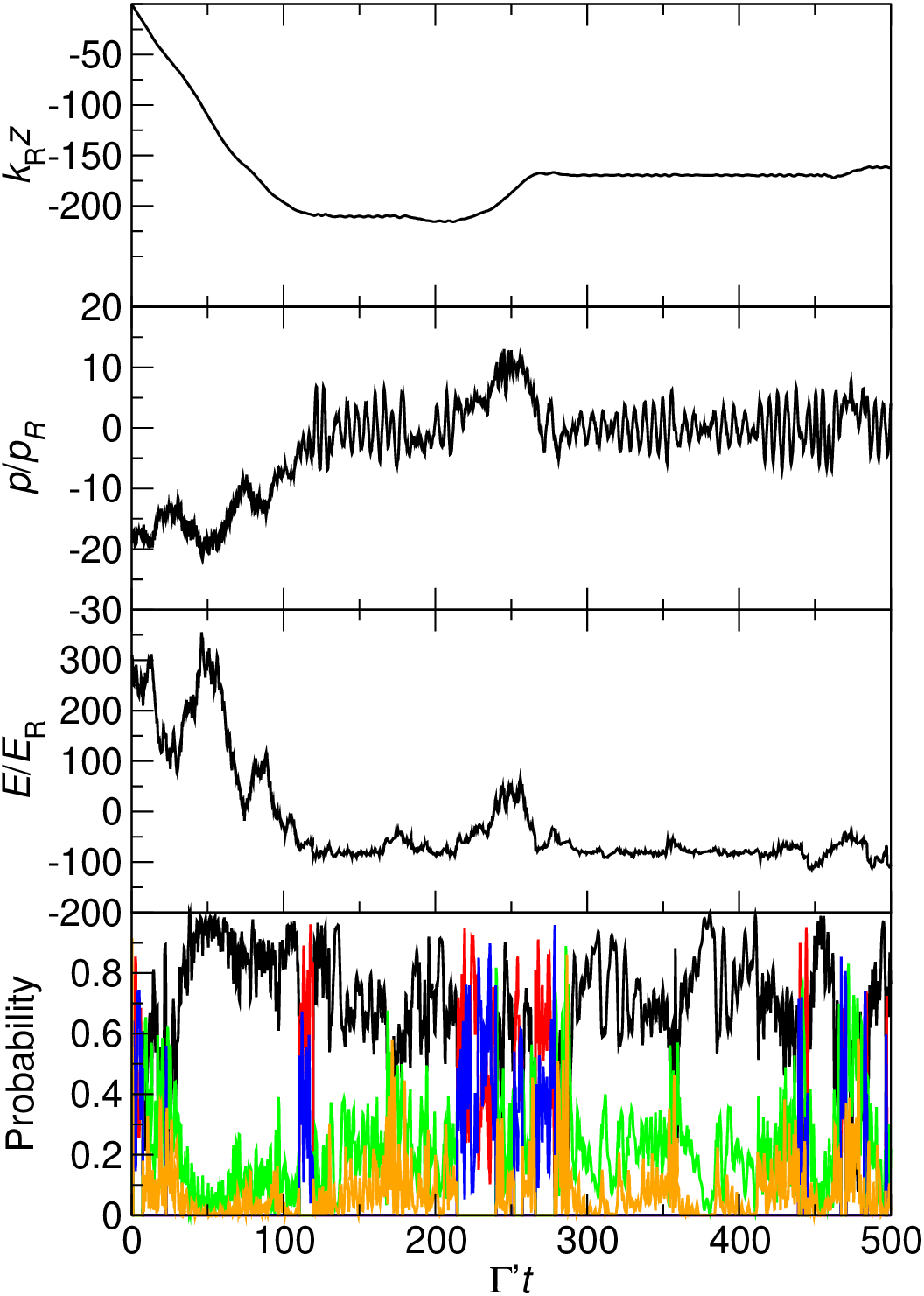}
  \caption{Position, momentum, energy, and internal state populations
    as a function of time for a single atom moving in an optical
    lattice. The potential depth is $\hbar|\Delta'|=150 E_{\mr R}$,
    and the detuning $\Delta=-10\Gamma$. The internal states have the
    color coding from Figure \protect\ref{fig:potentials}.}
  \label{fig:sa150}
\end{figure}

\begin{figure}[tbp]
  \includegraphics*[width=9cm]{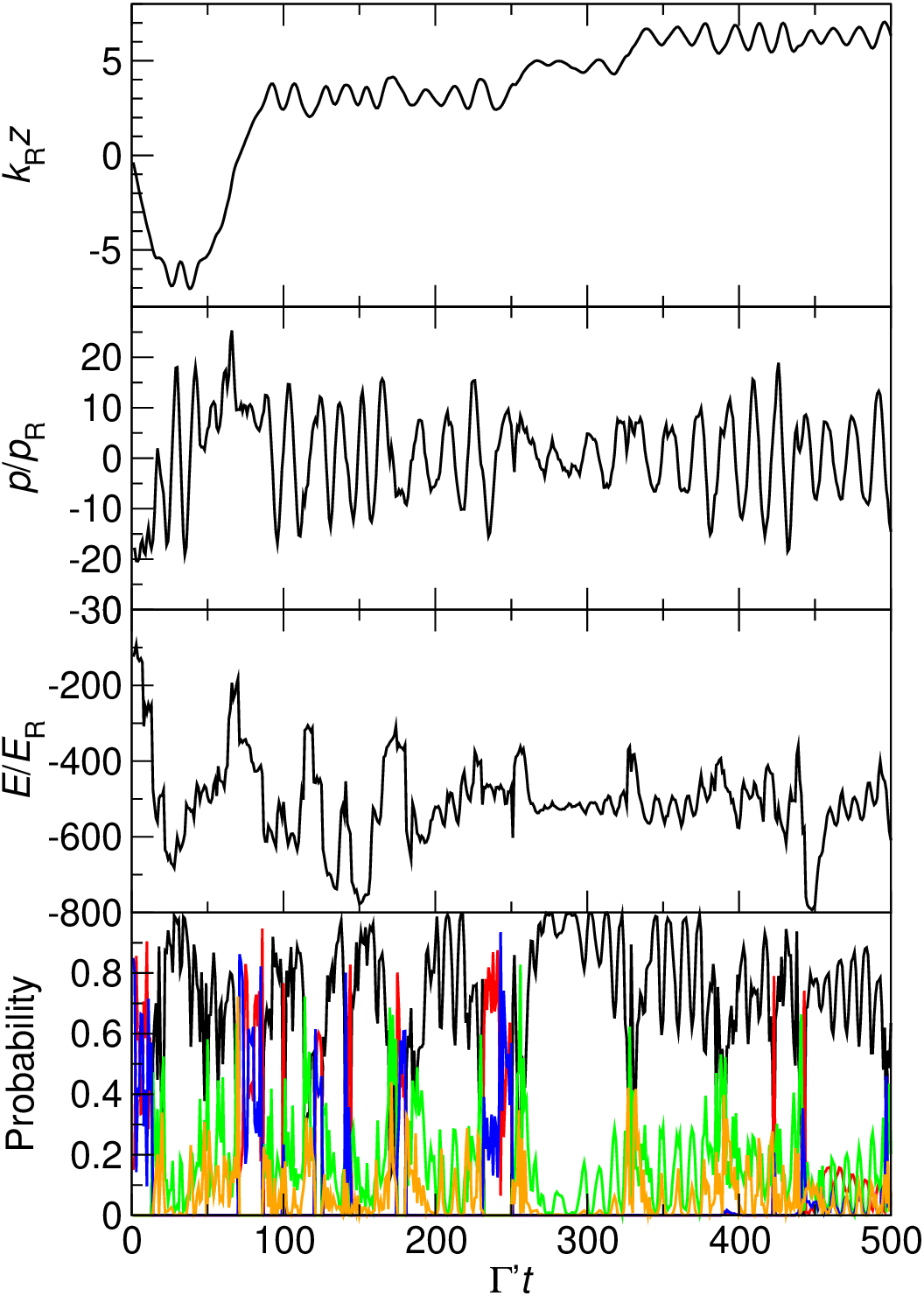}
  \caption{Same as Figure \ref{fig:sa150} but for a deeper potential,
    $\hbar|\Delta'|=1000 E_{\mr R}$ .}
  \label{fig:sa1000}
\end{figure}

The semi-classical method also makes it possible to follow the motion
of a single atom as it moves through the lattice.
In Figures \ref{fig:sa150} and \ref{fig:sa1000} we show the position,
momentum, energy and internal state distribution as a function of time
for a single atom in optical lattices with detunings
$\Delta=-10\Gamma$, and potential depths $\hbar|\Delta'|=150 E_{\mr
  R}$ and $\hbar|\Delta'|=1000 E_{\mr R}$ respectively. The energy was
calculated as the sum of the kinetic energy and light-shift potential,
i.e.,
\begin{equation}
  \label{eq:energy}
  E=\frac{p^2}{2m}+\hbar\Delta'{\rm Tr}\{A(z)w(t)\}.
\end{equation}

The ratio between the potential, pumping and diffusion terms in Eq.\ 
(\ref{eq:fullsc}) depends on $\Delta/\Gamma$ only, and is hence the
same in both graphs. The only difference lies in the inertial term
$p/m \partial_i$. Increasing $|\Delta'|$, while keeping the ratio
$\Delta/\Gamma$ constant, is equivalent to increasing the mass $m$ by
the same factor. This can be seen comparing the graphs, since the atom
is less mobile in Figure \ref{fig:sa1000}.

At both potential depths the atom shows, after an initial cooling
phase, a high degree of localization. While localized the atom
populates mostly the extreme magnetic states $M_{\mr g}=\pm J_{\mr
  g}$.  The energy is more or less constant, fluctuating around half
the potential depth. The amplitudes of the oscillations in momentum
and position vary somewhat due to diffusion, but tend to stay within
certain bounds as long as the atom remains in the same potential well.
We cannot see any clear trend towards smaller oscillation amplitudes
while the atom remains trapped in a site, i.e., we see no local
cooling.

The periods of localization are interrupted by brief phases where the
atom acquires enough energy to travel over many potential wells,
before once again getting localized.  These excursions are most
prominent at lower potential depths.  The periods when the atom is
untrapped are associated with abrupt changes of the internal state of
the atom, usually from odd to even magnetic states. (The light-shift
potential only induces odd--odd and even--even couplings between
magnetic states. Thus any pure quantum mechanical state is a
superposition of only odd or only even magnetic states.) During all
periods of localization the atom is in a state with similar
internal-state distribution and energy. Even when the energy sometimes
drops below this stationary value the atom is soon returned to the
same state.

These results are in qualitative agreement with our earlier conclusion
that Sisyphus cooling, especially at low potential depths, works
through a transfer of atoms between a hot and a cold mode
\cite{cla05}. The cold mode has a momentum distribution, with a width
that does not change over time. This mode corresponds to the
population of atoms in the trapped state. The cooling process is in
effect a transfer of atoms from the untrapped to the trapped state.

\begin{figure}[htbp]
  \includegraphics*[width=9cm]{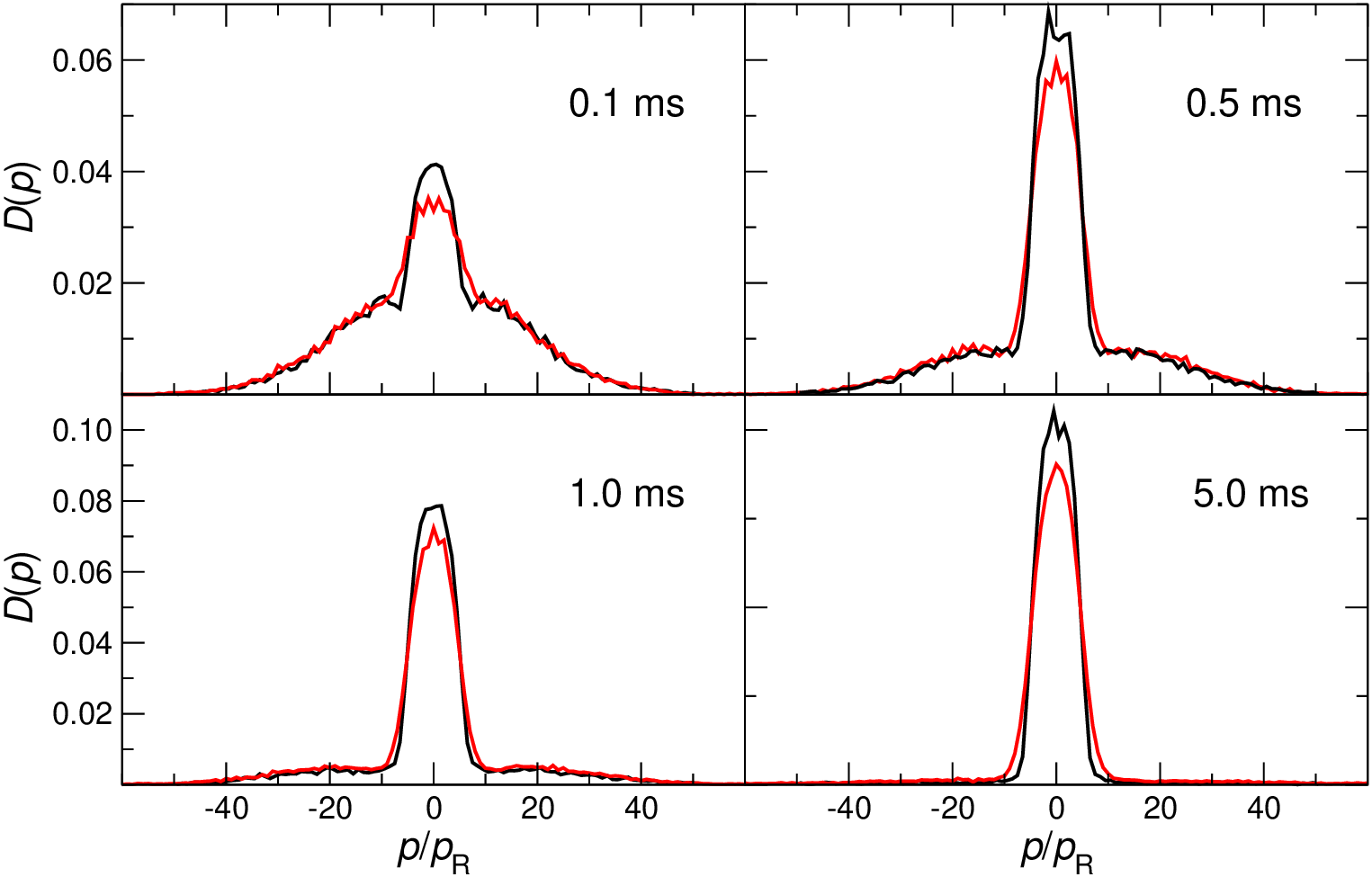}
  \caption{Time evolution of the momentum distribution for a potential
    depth $|\Delta'|=130 E_{\mr R}$. The starting temperature was $50
    \mu{\rm K}$. The black curve shows the semi-classical results,
    while the red curve shows results of a fully quantum mechanical
    simulation.}
  \label{fig:time}
\end{figure}

In Fig.\ \ref{fig:time} we compare the semi-classical approximation to
the time evolution of the momentum distribution $D(p)={\mr d}N(p)/{\mr
  d}p$ (where $N(p)$ is the number of atoms with momentum $p$) to the
results in \cite{cla05}, for $|\Delta'|=130 E_{\mr R}$. The bimodality
of the distribution is very clear also in the semi-classical results,
and the agreement with the quantum-mechanical results is very good.
The distribution of the hot mode is identical to within statistical
uncertainties. This shows that the physics of untrapped atoms,
including their rate of transfer to trapped states, is well described
by our semi-classical method. The semi-classical method gives a
slightly more narrow cold mode, in agreement with the results in Fig.\ 
\ref{fig:p2}.

\section{Discussion}

We have developed a semi-classical method to simulate the dynamics of
atoms in optical lattices.  Our results for the average momentum
distribution  of the atoms, including its time dependence, agree
excellently with those of the fully quantum mechanical method.
To achieve an accurate
description it is necessary to include both populations of and
coherences between the internal states of the atom. The external
degrees of freedom may, at least in some situations, be described
classically, i.e., as particles with definite positions and momenta.

The semi-classical approximation was introduced as a second order
Taylor expansion in $p/p_{\mr R}$ of the Wigner function. According to
our results $\langle p\rangle_{\text{rms}}\gtrsim 4 p_{\mr R}$, and
hence this expansion should be a fairly good approximation.
Nevertheless, there are some situations where the semi-classical
description must necessarily break down.  One is when effects from the
quantization of bound states are important. Such effects will be most
prominent when the atoms are localized near the bottom of the
potential wells. Another is the coherent splitting of a wave packet.
If the atomic wavefunction is, e.g.,  partially transmitted to the next
potential well, the semi-classical method will describe this as a
classical probability (some atoms are transmitted, some are not),
while any coherence effects between the two parts of the wave packet
will be lost.

The conceptual simplicity of the semi-classical descriptions makes it
a useful aid in visualizing complex physical processes.  It is also a
flexible tool, which is relatively easy to adapt to different physical
situations. In the near future we plan to extend the method to double
optical lattices \cite{ell03}. Further studies of the cooling process,
e.g.\ to deepen the understanding of the bimodal velocity
distributions observed in experiment and full quantum simulations, are
underway.


\section*{Acknowledgments}

We thank Robin Kaiser for useful discussions. This work was supported
by the Swedish Research Council (VR), Carl Tryggers stiftelse, and
Kempe stiftelserna. Part of the calculations were performed using the
resources of the High Performance Computing Center North (HPC2N).


\end{document}